\documentclass[
12pt
]{iopart}

\usepackage{xspace,amsfonts,amssymb,graphicx,color}
\begin{document}

\newcommand{\beq}{\begin{equation}}
\newcommand{\eeq}{\end{equation}}
\newcommand{\bqa}{\begin{eqnarray}}
\newcommand{\eqa}{\end{eqnarray}}
\newcommand{\nn}{\nonumber}
\newcommand{\nl}[1]{\nn \\ && {#1}\,}
\newcommand{\erf}[1]{Eq.~(\ref{#1})}
\newcommand{\erfs}[2]{Eqs.~(\ref{#1})--(\ref{#2})}
\newcommand{\dg}{^\dagger}
\newcommand{\rt}[1]{\sqrt{#1}\,}
\newcommand{\smallfrac}[2]{\mbox{$\frac{#1}{#2}$}}
\newcommand{\half}{\smallfrac{1}{2}}
\newcommand{\bra}[1]{\langle{#1}|}
\newcommand{\ket}[1]{|{#1}\rangle}
\newcommand{\ip}[2]{\langle{#1}|{#2}\rangle}
\newcommand{\sch}{Schr\"odinger }
\newcommand{\schs}{Schr\"odinger's }
\newcommand{\hei}{Heisenberg}
\newcommand{\heis}{Heisenberg's }
\newcommand{\bl}{{\bigl(}}
\renewcommand{\br}{{\bigr)}}
\newcommand{\ito}{It\^o }
\newcommand{\str}{Stratonovich }
\newcommand{\dbd}[1]{{\partial}/{\partial {#1}}}
\newcommand{\sq}[1]{\left[ {#1} \right]}
\newcommand{\cu}[1]{\left\{ {#1} \right\}}
\newcommand{\ro}[1]{\left( {#1} \right)}
\newcommand{\an}[1]{\left\langle{#1}\right\rangle}
\newcommand{\st}[1]{\left|{#1}\right|}
\newcommand{\implies}{\Longrightarrow}
\renewcommand{\tr}[1]{{\rm Tr}\sq{ {#1} }}
\newcommand{\del}{\nabla}
\newcommand{\du}{\partial}
\newcommand{\singlecol}{\end{multicols}
     \vspace{-0.5cm}\noindent\rule{0.5\textwidth}{0.4pt}\rule{0.4pt}
     {\baselineskip}\widetext }
\newcommand{\doublecol}{\noindent\hspace{0.5\textwidth}
     \rule{0.4pt}{\baselineskip}\rule[\baselineskip]
     
{0.5\textwidth}{0.4pt}\vspace{-0.5cm}\begin{multicols}{2}\noindent}
\newcommand{\tick}{$\sqrt{\phantom{I_{I}\hspace{-2.1ex}}}$}
\newcommand{\cross}{$\times$}
\newcommand{\ww}{which-way }
\newcommand{\ps}[1]{\hspace{-5ex}{\phantom{\an{X_{w}}}}_{#1}\!}

\newcommand{\bbb}[1]{\hspace{-5ex}{\phantom{\an{X_{w}}}}_{#1}\!}

\newcommand{\mket}[1]{\ket{#1}\hspace{-0.5ex}\rangle}
\newcommand{\mbra}[1]{\langle\hspace{-0.5ex}\bra{#1}}

%


\title{Adaptive Phase Measurements in Linear Optical Quantum 
Computation} 
\author{T. C. Ralph\dag\footnote[3]{email:ralph@physics.uq.edu.au}, A. P. Lund\dag\ and H. M. Wiseman\ddag}

\address{\dag\ Centre for Quantum Computer Technology,
Department of Physics, The University of Queensland,
St Lucia QLD 4072 Australia}

\address{\ddag\ Centre for Quantum Computer Technology, 
School of Science, Griffith University, Australia}

\begin{abstract}
Photon counting induces an effective nonlinear optical phase shift 
on certain states derived by linear optics from single photons. Although this nonlinearity 
is nondeterministic, it is sufficient in principle to allow scalable linear optics quantum computation (LOQC). The most obvious way to encode a qubit optically is as a superposition of the vacuum and a single photon in one mode --- so-called ``single-rail" logic. Until now this approach was 
thought to be prohibitively expensive (in resources) compared to ``dual-rail" logic where
a qubit is stored by a photon across two modes. Here we attack this problem with  
real-time feedback control, which can realize a quantum-limited phase measurement 
on a single mode, as has been recently demonstrated experimentally. We show that with this added measurement resource, the resource requirements for single-rail LOQC are not substantially different from those of dual-rail LOQC. In particular, with adaptive phase measurements an arbitrary qubit state $\alpha \ket{0} + \beta\ket{1}$ can be prepared {\em deterministically}. 
\end{abstract}


\maketitle

\section{Introduction}

Linear optical quantum computation (LOQC) employs linear networks to process
encoded photonic qubits.  The necessary non-linearities are induced via photon
number resolving measurements which  post-select  desired outcomes.  In
principle, the results of these detections can be fed-forward  in  the network to
produce a scaleable system.  The first proposal of this type was made by Knill,
Laflamme and Milburn (KLM)  \cite{knill}. Since then a number of  variations
have been discussed \cite{lund,franson,ralph,nielsen}. 

At the basic level the scheme is non-deterministic. A number of proof of
principle experiments have now been performed at this level
\cite{pit03,job03,gas04,zha05}.  Scale-up relies on implementing particular
types of ``teleported'' gates with a specific failure mode (qubit measurement)
which can be encoded against in various ways \cite{knill,nielsen,hayes}.

One point in common of all photonic LOQC proposals to date is that they
exclusively employ photon counting for all measurements. It is of interest to
investigate if other types of measurements can also play a role.  Here we show
that adaptive phase measurements \cite{Wis95c} can be used to accomplish some
useful tasks in LOQC schemes.  Adaptive phase measurements are based on
homodyne detection and thus  the measurement outcome has a continuous 
spectrum --- quite
unlike  the discrete values produced by a photon counter. 

 In this paper we  will focus on single photon, ``single-rail'' schemes, 
 where phase-sensitive optical detection has the most obvious application. 
  In these schemes the
logical values of the optical qubits correspond to the occupation numbers of
their respective field modes, i.e.: $|0 \rangle_{L} = |0 \rangle$ and $|1
\rangle_{L} = |1 \rangle$ where $|n \rangle$ is a single mode, $n$ photon Fock
state. In contrast ``dual-rail'' schemes encode qubits between two modes as $|0
\rangle_{L} = |0 1 \rangle$ and $|1 \rangle_{L} = |1 0 \rangle$ where $\ket{n
m}$ is the direct product of two single mode Fock states,  with occupation
numbers of $n$ and $m$ respectively.  Basic non-deterministic gate operations
have been described using single rail encoding \cite{lund} and experimental
demonstrations of some basic operations have been made
\cite{lom02,lvo02,bab04}. 


 Before introducing adaptive measurements,  
 it is useful to compare the two schemes, single-rail and dual-rail, when only photon counters are used (see Table~\ref{comparison_before}). Dual-rail requires single photons as resouces and photon counters for detection.  Single-rail requires the same plus a supply of coherent states.  Preparing arbitrary initial states and performing single qubit gates in dual-rail can be achieved with unit
probability. However, in single-rail these operations are non-deterministic with low probability of success
and require complicated interferometers.  The two-qubit control sign gate is non-deterministic and of equal difficulty in both schemes. Because of the difficulty in performing single qubit operations this comparison suggests that single-rail schemes are much more difficult to implement than dual-rail schemes. We will revisit this comparison in the conclusion after adding adaptive phase measurements to our single-rail tool-kit. Before leaving the comparison it is worth noting the distinct effect of photon loss in each scheme. Photon loss in the dual-rail scheme moves the qubit outside
the logical basis whilst in the single-rail scheme photon loss can be treated as a
logical error. This means that standard error correction schemes could be used
in a single-rail scheme to  combat loss and other logical errors simultaneously. In dual-rail schemes a separate protocol targeted specifically at loss errors would be required. That said we note that an efficient coding against loss in dual rail has been described \cite{ral05}.

\begin{table}[htb]
\begin{center}
  \begin{tabular}{||r||c|c||}
\hline
   & Dual-Rail & Single-Rail \\ \hline\hline
Qubit  & $c_0\ket{01} + c_1\ket{10}$  & $c_0\ket{0} + c_1\ket{1}$ \\  \hline 
State resources & $\ket{1}$ & $\ket{1}$, $\ket{\alpha}$ \\ \hline
Detector resources & photon counters & photon counters \\ \hline
Preparing a qubit  & deterministic, free & {\em prob.}, {\em expensive
} \\ \hline
Single qubit gates & deterministic, free & {\em prob.}, {\em expensive
}  \\ \hline
Two-qubit gates & prob., cheap & prob., cheap \\ \hline 
Photon loss & {\em Outside encoding} & Logical error \\  
\hline
$\implies$ Q. Encoding & {\em beyond standard} & standard \\ \hline
 \hline
\end{tabular} 
\end{center}
\caption{\label{comparison_before}Comparison of the dual-rail and single-rail
encodings prior to this work. Here `prob.' stands for probabilistic and means
can be done with probability less than unity. `Free' means can be achieved with
only one use of the resource to achieve a unit fidelity,  `cheap'  means can be
achieved with a finite number of resources  (on average)  to achieve unit fidelity and
 `expensive'  means unit fidelity can only be achieved asymptotically, with
arbitrarily many resources.  $\ket{0}$ and $\ket{1}$ are Fock states
and $\ket{\alpha}$ represents a coherent state.  Italics are used to indicate 
apparent disadvantages.  } 
\end{table}


The paper is arranged in the following way. In the next section we will
describe adaptive phase measurements and derive the relevant projector
appropriate for single rail encoding measurements. In section 3 we will
describe how a single rail superposition state can be deterministically
produced from single photon number states using adaptive phase measurements. In
section 4 we will consider another application; the reduction of dual-rail
encoding to single rail encoding using adaptive phase measurements and hence
describe how arbitrary rotations of single rail qubits can be implemented based
on this technique. Finally in section 5 we will conclude.

\section{The Adaptive Phase Measurement}

In this section we show how the required phase measurements can be achieved using 
adaptive dyne detection. Here `dyne' detection means detection using a large local oscillator in a coherent state, 
as in heterodyne and homodyne detection. This can be achieved with high efficiency \cite{bachor} 
and we will assume an efficiency of unity. Consider a single mode (i.e. transformed-limited)  pulse 
with an envelope $u(t) \geq 0$. This pulse shape $u(t)$  is assumed to be $0$ for $t<0$ and $t>T$ and 
to be normalized so that $U(T)=1$, where
\beq
U(t) = \int_0^t u(s) ds
\eeq
We assume a mode-matched local oscillator, that is, with the same pulse shape, but we note that in principle a CW local oscillator could be used, and the dyne current corrected electronically by multiplying it by the factor $\sqrt{u(t)}$.  
Let the phase of the local oscillator be $\Phi(t)$, which is fixed for homodyne detection and 
rapidly varying (linearly in time) for heterodyne detection.  In the large local oscillator limit, the dyne current for times $0<t<T$ is given by
\beq \label{I}
 {I}(t) = e^{-i\Phi(t)}b(t) + {\rm H.c.}.
\eeq
Here $ {b}(t)$ is proportional to the annihilation operator for the field incident upon the detector at time $t$,  and is  given by 
\beq \label{b}
b(t) = u(t)   a - u(t)\int_0^t \frac{\rt{u(s)} {\nu}(s)}{1-U(s)} ds + \rt{u(t)} \nu(t) 
\eeq
Here $ {a}$ is the annihilation operator for the pulse, while $ {\nu}(t)$ is the annihilation operator for the localized vacuum incident upon the detector at time $t$. It obeys
\beq
 [ {\nu}(t), {\nu}(s)\dg] = \delta(t-s).
\eeq
Note that in the mean we have simply $\an{b(t)} = u(t)\an{a}$.
The equation (\ref{b}) can be derived using the results of Ref.~\cite{Wis95qo1}, generalizing the mode shape $\gamma e^{-\gamma t}$ of a freely decaying cavity to an arbitrary mode shape $u(t)$ by defining a time-dependent cavity decay rate $\gamma(t) = -(d/dt) \ln [1-U(t)]$. 

Before considering adaptive dyne measurements, it is instructive to consider conventional dyne detection, such as homodyne detection. In this case $\Phi$ is a constant, and for a single-mode pulse we are interested only in the integrated current
\beq \label{X1}
 {X} = \int_0^T  {I}(t) dt.
\eeq
Using integration by parts [without relying on any properties of $\nu(t)$], 
this can be evaulated from \erf{I} to be 
\beq \label{X2}
 {X} =  {a}e^{-i\Phi} +  {a}\dg e^{i\Phi},
\eeq
 the $\Phi$-quadrature of the pulse.  
What \erf{X2} means is that the probability density for the integrated current is 
\beq \label{px}
\wp(X=x)dx  = \bra{x}\rho\ket{x} dx,
\eeq
where $\ket{x}$ is an eigenstate of $ {a}e^{-i\Phi} +  {a}\dg e^{i\Phi}$ normalized so that $\ip{x}{x'}=\delta(x-x')$ 
and $\rho$ is the system state. 

In  {\em adaptive} dyne detection, the phase $\Phi(t)$ is made to depend
upon $  I(s)$ for $s<t$. This was first treated quantitatively in Ref.~\cite{Wis95c}. 
It requires a real-time feedback loop with a time delay 
much smaller than the characteristic width of the pulse, as realized 
for the first time in Ref.~\cite{Arm02}. This leads to a nonlinear, non-Markovian 
equation for $ {I}(t)$, and the problem is intractable in general.  However, for the particular 
case of the following adaptive algorithm 
\beq \label{Phi}
\Phi(t) = \int_0^t \frac{I(s)ds}{\sqrt{U(s)}},
\eeq
introduced in Ref.~\cite{Wis95c} and implemented in Ref.~\cite{Arm02}, 
the problem can be solved semi-analytically \cite{WisKil98} using the theory of generalized quantum measurement.  Moreover, for the particular case where the state contains at most one photon then
there is an analytical solution  \cite{Wis95c}  involving a single integral of the current, just as in the example of homodyne detection above. It turns out that the integral is the the same as that used in the feedback (\ref{Phi}):
\beq
\Theta =  \int_0^T \frac{I(t)dt}{\sqrt{U(t)}} -  \frac{\pi}{2}.  
\eeq
Unlike $X$, $\Theta$ cannot be represented by an operator in the system Hilbert space. But the equation analogous to \erf{px} is simply
\beq
\wp(\Theta=\theta) d\theta = \bra{\theta}  \rho \ket{\theta} d\theta /  2\pi,
\label{proj}
\eeq
where 
\beq
\ket{\theta} = \ket{0} + e^{i\theta}\ket{1}.
\eeq

This adaptive dyne detection is useful for estimating the unknown phase of an optical pulse. But at the single-photon level we are more interested in its ability in state preparation. Say the system is entangled with other modes, so that the total state is $\mket{\Psi}$. Then the conditioned state of the other modes after the measurement  yielding result $\theta$ is 
\beq \label{prep}
\langle{\theta}\mket{\Psi}/\sqrt{2\pi},
\eeq 
where the squared norm of this state is equal to the probability density for obtaining this outcome.  
We discuss such applications in the next section.

\section{Superposition State Preparation}

A basic qubit operation is the production of superposition states. For
dual-rail systems arbitrary superposition states can be prepared using only
linear optics, provided single photon states of a single mode can be prepared.
Arbitrary single-rail superposition states, $\alpha |0 \rangle + e^{-i
\phi}\sqrt{1-\alpha^{2}} |1 \rangle$, with $\alpha$ and $\phi$ real
numbers, are not so easy to produce.  Previous suggestions for deterministic production of such states involved non-linearities significantly larger than currently feasible. Alternatively, non-deterministic techniques based on photon counting
\cite{pegg,lund} have been described and experimentally demonstrated  \cite{lvo02} but these have low probabilities of success. A non-deterministic scheme based on homodyne detection has also been demonstrated \cite{bab04}. We now show that it is possible to deterministically
produce an arbitrary single-rail state from a single photon state using linear
optics and adaptive phase measurements.

Consider first producing an equal superposition state. We begin by splitting a single photon equally into two modes, so that $\mket{\Psi}=(\ket{0,1}+\ket{1,0})/\sqrt{2}$. If we then do an adaptive phase measurement on the first mode we obtain a result $\theta$, which,  from \erf{proj}, is drawn randomly from $[0,2\pi)$. From \erf{prep},  the second photon is prepared in state $\ket{-\theta}$. Then by feedforward onto an phase modulator,  this random phase can be removed, yielding deterministically the state
\beq
(\ket{0}+\ket{1})/\sqrt{2}.
\eeq
This can be contrasted with the technique used in Ref.~\cite{bab04} that relied upon homodyne detection. There the second mode is prepared in a state with known phase, but with a random amplitude:
\beq
(x\ket{0} + \ket{1} )/\sqrt{1+x^2}
\eeq
Here from \erf{px} the probability distribution for $x$ is 
\beq
\wp(x)dx= e^{-x^2/2} \frac{1+x^2}{ 2\sqrt{2\pi}} dx.
\eeq
The probability for obtaining $x=1$ is infinitesimal. A finite probability of success necessarily leads to a drop in fidelity.

Our procedure is easily generalized to the production of an arbitrary superposition state. The  single photon state is now mixed with a vacuum
state on a beamsplitter with intensity reflectivity $\eta$.  The resulting
evolution is
\begin{equation}
    |1 \rangle |0 \rangle \to \sqrt{\eta} |1 \rangle |0 \rangle + 
    \sqrt{1 - \eta} |0 \rangle |1 \rangle
    \label{sr1}
\end{equation}
We then make an adaptive phase meansurement on the reflected output port and
introduce a phase delay $\psi$ to the transmitted mode.  If the result of the
adaptive phase measurement is $\theta$, the conditional state of the transmitted
mode  is, from \erf{prep}, 
\begin{equation}
    \sqrt{\eta}  |0 \rangle + e^{-i (\theta + \psi)}
    \sqrt{1 - \eta} |1 \rangle. 
    \label{sr2}
\end{equation}
 By choosing $\sqrt{\eta} = \alpha$ and $\psi = \phi-\theta$ an arbitrary state can be produced deterministically.

\section{Quantum Gates using Adaptive Phase Measurements}

We have shown that arbitrary single rail, single qubit states can be produced
deterministically from single photon states using adaptive phase measurements. For quantum processing
however we require the ability to impose arbitrary rotations on unknown input
states. This is a more difficult problem as it requires a Hadamard gate defined
by the transformations  $|0 \rangle \to (|0 \rangle + |1
\rangle)/\sqrt{2}$ and $|1 \rangle \to (|0 \rangle - |1 \rangle)/\sqrt{2}$.  A
Hadamard transformation plus the phase rotations that come from propagation
allow arbitrary rotations. A non-deterministic Hadamard transformation for
single rail qubits based on photon counting has previously been described
\cite{lund}. However, the success probability of this construction  was very
low. Here we show that a  Hadamard transformation based on a combination of
photon counting and adaptive phase measurements, whilst still non-deterministic, can be much more
efficient.
 
The key observation underlying this new construction is that a deterministic
mapping of dual rail encoding into single rail encoding can be achieved using
adaptive phase measurements. Consider the arbitrary dual rail qubit $\alpha |0 1 \rangle + e^{-i
\phi}\sqrt{1-\alpha^{2}} |1 0 \rangle$. Suppose an adaptive phase measurement is made on the second
rail of the qubit giving the result $\theta$ and a phase delay of $-\theta$ is
subsequently imposed on the remaining rail of the qubit. The resulting state of
the remaining rail is $\alpha |0 \rangle + e^{-i \phi}\sqrt{1-\alpha^{2}} |1
\rangle$, which is a single rail qubit with the same logical value as the
original dual rail qubit. 

What about the reverse operation? Is it possible to go from a single rail
encoded qubit to a dual rail encoding? It does not appear to be possible to do
this deterministically with only linear  optics. However  a non-deterministic
transformation is possible using teleportation. The idea is to teleport between
encodings. Dual rail teleportation can be achieved using the dual rail Bell
state $\ket{01} \ket{10} + \ket{10} \ket{01}$.  (We are ignoring normalization for 
convenience.)  Single rail teleportation can be
achieved using the single rail Bell state $\ket{0} \ket{1} + \ket{1} \ket{0}$.
In both cases only two of the four Bell states can be identified with linear
optics so the teleportation works 50\% of the time. Now suppose we take a dual
rail Bell state and use an adaptive phase measurement to project one of its arms into a single rail
state. We obtain the state $\ket{0} \ket{10} + \ket{1} \ket{01}$ which is Bell
entanglement between dual and single rail qubits.  If we now perform a Bell
measurement between the single rail half of the entanglement and  an arbitrary
single rail qubit then (when successful) the qubit will be teleported onto the
dual rail part of the entanglement, thus converting a single rail qubit into a
dual rail qubit.

We now have a way of (non-deterministically) performing an arbitrary rotation
on an unknown single rail qubit. The idea is depicted schematically in Fig.1.
First we teleport the single rail qubit onto a dual rail qubit. Then we perform
an arbitrary rotation on the dual rail qubit. We then use an adaptive phase measurement to transform the
dual rail qubit back to a single rail qubit. The only non-deterministic step is
the Bell measurement in the teleportation thus the arbitrary rotation will
succeed 50\% of the time. This is a major improvement in success probability
over previous schemes.

The fundamental 2 qubit operation, the control-sign gate, used in dual
rail implementations is in fact a single rail gate \cite{lund2}. The
combination of arbitrary single qubit rotations and the control-sign gate form a universal set. 
\begin{figure}
\begin{center}
\includegraphics[width=12cm]{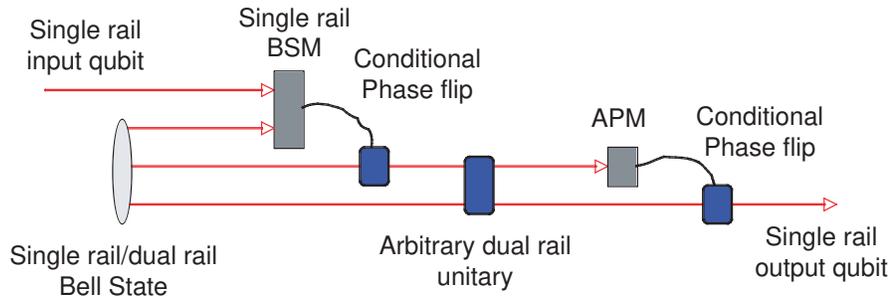}
\caption{A schematic representation of the application of an arbitrary single qubit unitary to a single rail qubit. The single rail qubit is teleported onto a dual rail qubit, the unitary is applied, then an adaptive phase measurement is used to convert back to a single rail qubit. BSM $\equiv$ Bell state measurement. APM $\equiv$ Adaptive phase measurement. All operations are deterministic except the Bell state measurement which succeeds 50\% of the time.}
\label{f1}
\end{center}
\end{figure}

\section{Conclusion}

We have shown that an adaptive phase measurement, when applied to an input state containing at most one photon, effectively measures the overlap of the input state with the measurement state $\ket{0} + e^{i\theta}\ket{1}$, where $\theta$ is the measurement outcome. We have applied this measurement technique to the deterministic preparation of arbitrary superposition states of single-rail photonic qubits. Previously it was thought that large non-linearities would be required for such operations. We have also applied the technique to the application of single qubit quantum gates for single rail qubits and described a more efficient (though still non-deterministic) protocol.  The increase in efficiency when using adapative phase measurements with single-rail qubits is summarised in Table~\ref{comparison_after} ({\em c.f.} Table~\ref{comparison_before}). 

To summarize, this work introduces a new tool to the LOQC tool-box and increases the practicality of demonstrating and using single rail photonic qubits. Other applications for real-time feedback control techniques in LOQC remain to be explored. 

\begin{table}[htb]
\begin{center}
  \begin{tabular}{||r||c|c||}
\hline
   & Dual-Rail & Single-Rail \\ \hline\hline
Qubit  & $c_0\ket{01} + c_1\ket{10}$  & $c_0\ket{0} + c_1\ket{1}$ \\  \hline
State resources & $\ket{1}$ & $\ket{1}$, $\ket{\alpha}$ \\ \hline
Detector resources & PC & PC, {\em APM} \\ \hline
Preparing a qubit  & deterministic, free & {\em deterministic, free} \\ \hline
Single qubit gates & deterministic, free & {\em prob., cheap}  \\ \hline
Two-qubit gates & prob., cheap & prob., cheap \\ \hline
Photon loss & {Outside encoding} & Logical error \\
\hline
$\implies$ Q. Encoding & {beyond standard} & standard \\ \hline
 \hline
\end{tabular}
\end{center}
\caption{\label{comparison_after}Comparison of the dual-rail and single-rail
encodings considering this work.  This table is the same format as
Table~\ref{comparison_before} but photon counters has been abbreivated to PC.
 This time, italics indicate changes from Table~\ref{comparison_before}: 
 including  adapative phase measurements
(APM) eliminates the resource problems in preparing an arbitrary qubit and
performing single qubits gates.}
\end{table}


\ack

This project was supported by the Australian Research Council and the Queensland State Government.

\newcommand{\pra}{Phys. Rev. A}
\newcommand{\prl}{Phys. Rev. Lett.}

%

\end{document}